\documentclass[twocolumn,floatfix,showpacs,prd,aps,tightenlines,superscriptaddress]{revtex4}
\usepackage{graphicx}
\usepackage{amssymb}
\usepackage{mathrsfs}

\usepackage{dcolumn}
\usepackage{bm}

\newcommand{\bea}{\begin{eqnarray}}
\newcommand{\eea}{\end{eqnarray}}
\newcommand{\beq}{\begin{equation}}
\newcommand{\eeq}{\end{equation}}
\newcommand{\KMS}{{\rm km\ s^{-1}}}

\newcommand{\scri}{\mathscr{I}}

\begin{document}

\title{Quasi-Local Linear Momentum in Black-Hole Binaries}

\author{Badri Krishnan}
\affiliation{Max-Planck-Institut f\"ur Gravitationsphysik,
Albert-Einstein-Institut, Am M\"uhlenberg 1, D-14476 Golm, Germany}

\author{Carlos O. Lousto}
\affiliation{Center for Computational Relativity and Gravitation,
School of Mathematical Sciences,
Rochester Institute of Technology, 78 Lomb Memorial Drive, Rochester,
 New York 14623}

\author{Yosef Zlochower}
\affiliation{Center for Computational Relativity and Gravitation,
School of Mathematical Sciences,
Rochester Institute of Technology, 78 Lomb Memorial Drive, Rochester,
 New York 14623}

\date{\today}

\begin{abstract}
  We propose a quasi-local formula for the linear momentum of
  black-hole horizons inspired by the formalism of quasi-local
  horizons.  We test this formula using two complementary
  configurations: (i) by calculating the large orbital linear momentum
  of the two black holes in an unequal-mass, zero-spin,
  quasi-circular binary and (ii) by calculating the very small recoil
  momentum imparted to the remnant of the head-on collision of an
  equal-mass, anti-aligned-spin binary.  We obtain results consistent
  with the horizon trajectory in the orbiting case, and consistent
  with the net radiated linear momentum for the much smaller head-on recoil
  velocity.
\end{abstract}

 \maketitle

\section{Introduction}\label{sec:introduction}

The major breakthroughs in numerical relativity that occurred two
years ago~\cite{Pretorius:2005gq,Campanelli:2005dd,Baker:2005vv} not
only allowed numerical relativists to faithfully simulate the
inspiral, merger, and ringdown of arbitrary
black-hole-binary configurations, but also inspired new developments
in mathematical and numerical relativity and astrophysics. These include
exploring
cosmic censorship~\cite{Campanelli:2006uy}, modeling the horizon spin
direction in non-axisymmetric spacetimes~\cite{Campanelli:2006fy},
modeling black-hole---neutron-star binaries~\cite{Shibata:2006ks},
modeling the collapse of supermassive stars~\cite{Liu:2007cf}, and
finding recoil velocities in astronomical
observations~\cite{Bogdanovic:2007hp,Loeb:2007wz,Bonning:2007vt}.
These breakthroughs have also had significant consequences for the data
analysis of LIGO and other ground based observatories that are
attempting to directly detect gravitational
waves~\cite{Porter:2007vk,Ajith:2007qp,Pan:2007nw}.

The recoil velocities acquired by the remnant of the merger of
black-hole binaries has many interesting astrophysical
consequences~\cite{Campanelli:2007ew}, particularly since spinning
black holes can accelerate the merged hole up to
$4000\KMS$~\cite{Campanelli:2007cg}, large enough to eject the remnant
from the host galaxy.  As was noted
in~\cite{Herrmann:2007ac,Campanelli:2007ew,Koppitz:2007ev}, the spin
contributions to the recoil velocity are generally larger than those
due to the unequal masses, and, in particular, the spin component
in the orbital plane has the largest effect~\cite{Campanelli:2007ew}, leading
to a maximum recoil velocity of about
$4000\KMS$~\cite{Campanelli:2007cg}.  The first
study of generic black-hole-binary configurations (i.e.\ a binaries
with unequal component masses and spins, and spins not aligned with
each other or the orbital angular momentum) was described
in Ref.~\cite{Campanelli:2007ew}, and, based on the results of that
study, a semi-empirical formula to estimate the recoil velocities of
the remnant black holes was proposed, finding recent confirmation
in~\cite{Campanelli:2007cg,Herrmann:2007ex,Bruegmann:2007zj}.

In all the above papers the computation of recoil velocities was made
by directly extracting the radiated linear momentum of the system at
large distances from the remnant black hole (asymptotically
approaching future null infinity $\scri^+$)~\cite{Campanelli99}. Here
we are interested in an alternative measure of linear momentum; one in
terms of quantities defined local to the horizon.  Such a formula
would provide an independent computation of recoil velocities (and
thus provides a further consistency check for numerical simulations)
and would allow for the computation of the linear momentum of the
individual holes while they orbit each other before the final plunge
and merger.  The starting point of our analysis is the quasi-local
description of black hole horizons as provided by the
isolated~\cite{Ashtekar:2000sz}, dynamical~\cite{Ashtekar:2002ag}, and
trapping horizon~\cite{Hayward:1993wb} formalisms; see
\cite{Ashtekar:2004cn,Gourgoulhon:2005ng,Booth:2005qc} for reviews.
These techniques have been successfully applied to black-hole binaries
to extract the mass and spins (magnitude and direction) of the
individual holes and the merger remnant, as well as measure the
precession of the spin direction and spin
flips~\cite{Dreyer:2002mx,Campanelli:2006fg,Campanelli:2006fy,Cook:2007wr}.
In this paper we propose, and numerically test, a formula inspired by
the theory of quasi-local horizons to compute the linear momentum of a
black hole.

\section{Quasi-local Linear Momentum}
\label{sec:momentum}

In standard classical mechanics and field theory, conserved quantities
are defined, using Noether's theorem, as the generators of symmetries
of the action.  Thus, angular momentum is defined as the generator of
rotational symmetry, energy is the generator of time translations, and
linear momentum the generator of spatial translations.  This approach
has been successfully used to calculate the energy and angular
momentum of isolated and dynamical horizons
\cite{Ashtekar:2000hw,Ashtekar:2001is,Booth:2001gx,Booth:2005ss}. For
classical mechanics and field theory in flat Minkowski space, there is
a natural way to define translations and rotations using the
symmetries of the flat background metric.  The situation is more
complicated for general relativity on a curved spacetime manifold when
we typically do not have the luxury of being able to use a reference
background metric; this is precisely the situation for numerical
relativity black-hole simulations.  To even speak about conserved
quantities in this context, one therefore needs to assume the
existence of an (at least approximate) rotational symmetry for angular
momentum, and similarly a preferred time translation vector field at
the horizon for energy.  Since the spacetime in the vicinity of a
black-hole binary is not translationally invariant, it is clear that
the Hamiltonian calculations cannot be simply extended to define
linear momentum for the individual black holes.  Because of this
conceptual difficulty, we follow a different approach for linear
momentum: we shall take the Hamiltonian calculations for angular
momentum and energy only as motivations for our definition of linear
momentum. As we shall see, while our results are very promising, there
are open issues remaining; this work should therefore be seen as a
preliminary investigation to be followed up with further mathematical
and numerical work.

Motivated by the analysis of angular momentum in
\cite{Ashtekar:2003hk}, let us start with the momentum constraint on a
spatial slice $\Sigma$ of the spacetime with extrinsic curvature
$K_{ab}$ and 3-metric $\gamma_{ab}$:
\begin{equation} 
D^aP_{ab} = 0 \qquad \textrm{where} \qquad  P_{ab} = K_{ab}-K\gamma_{ab}\,. 
\end{equation}
Here $D_a$ is the derivative operator on $\Sigma$ compatible with
$\gamma_{ab}$.  Let $\xi^a$ be a vector field tangent to $\Sigma$
which looks asymptotically like a unit translation, and $\gamma_{ab}$
is the 3-metric.  Let us first consider the case when there is a
single apparent horizon $S$, considered to be the inner boundary of
$\Sigma$, and let $S_\infty$ be the sphere at spatial infinity.  We
are interested in the linear momentum along $\xi^a$.  Contracting the
momentum constraint with $\xi^a$ and integrating by parts, we get
\begin{equation} 
  \label{eq:fluxmom} 
  \oint_{S} - \oint_{S_\infty}{P}_{ab}\xi^a \, d^2S^b = \frac{1}{2}
  \int_\Sigma P^{ab}\mathcal{L}_\xi {\gamma}_{ab}\, d^3V\, , 
\end{equation}
where $\mathcal{L}_\xi$ is the Lie derivative with respect to $\xi^a$.
We have assumed asymptotic flatness at spatial infinity with $K_{ab} =
\mathcal{O}(r^{-2})$ and $\gamma_{ab} = \delta_{ab} +
\mathcal{O}(r^{-1})$ at large distances from the source.  Since our
focus here is near the horizon, we shall not spell out any further
details of the asymptotic behavior.  

We would like to interpret Eq.~(\ref{eq:fluxmom}) as a balance law for
linear momentum.  It should be noted that the analog of this equation
for a rotation $\xi^a=\varphi^s$ leads to the standard expressions for
angular momentum at infinity and at the
horizon~\cite{Ashtekar:2003hk}. As expected, the surface integral at
infinity is the usual ADM momentum associated with the entire Cauchy
surface, and we would like to identify the surface term at $S$ with
the linear momentum of the black hole:
\begin{equation} 
  \label{eq:Pxi}
  P_\xi^{(S)} = \frac{1}{8\pi}\oint_{S}\left({K}_{ab}-K\gamma_{ab}\right) \xi^a \,
  d^2S^b\,.
\end{equation}
The factor of $8\pi$ is chosen to recover the ADM linear momentum at
$S_\infty$ (and we have chosen units with $G=1$).  The right had side
of Eq. (\ref{eq:fluxmom}) is interpreted (up to a factor of $8\pi$) as
the ``flux'' of linear momentum between the horizon and infinity; it
would vanish if $\xi^a$ is a Killing vector of the 3-metric.  We do
not expect this flux term to vanish generically. However, it does
vanish for maximal slices ($K=0$) if $\xi^a$ is a conformal Killing
vector of the 3-metric $\gamma_{ab}$.  In particular, the flux
vanishes for the conformally flat Bowen-York data for a single boosted
black hole if we take $\xi^a$ to be any of the coordinate basis
vectors $\partial/\partial x,\partial/\partial y,\partial/\partial z$.
Thus, in this case $P_\xi$ agrees with the ADM linear momentum.  For
Bowen-York data with $N$ multiple boosted spinning black holes, if we
denote the $i^{th}$ apparent horizon as $S^i$, the same argument as
above shows that $P_{\xi}^{(S^i)}$ is additive:
\begin{equation}
  \label{eq:sum-BYmom}
  P_\xi^{ADM} = \sum_{i=1}^N P_{\xi}^{(S^i)}\,.
\end{equation}
Thus we see that, for Bowen-York initial data and, somewhat more
generally, for maximal conformally flat spatial slices,
Eq.~(\ref{eq:Pxi}) can be considered as a satisfactory formula for
quasi-local black hole linear momentum.  Moreover, the derivation of
the balance law Eq. (\ref{eq:fluxmom}) did not anywhere use the fact
that $S$ is a marginally trapped surface; the balance law is valid for
any inner boundary.  In particular then for maximal conformally flat
slices, $P_{\xi}^{(S)}$ can be used to measure the linear momentum
contained in closed regions in $\Sigma$.

In the general case during an evolution, we will have neither
conformal flatness nor maximal slices. The linear momentum flux will
not vanish, and $P_\xi^{(S)}$ will not agree with $P_\xi^{ADM}$.  This
fact, by itself is not necessarily a problem.  The most obvious
problem with interpreting $P_\xi$ as linear momentum then is the
choice of $\xi^a$ at the horizon; what should it's direction be, and
how should it be normalized?  In the regime when the black holes are
sufficiently far apart and the orbit is varying slowly, a possible
choice for $\xi^a$ is (spatial projection of) the helical vector
tangent to the orbit.  Alternatively, the correct method might be to
find an approximate translation Killing vector field on $\Sigma$ in
the vicinity of the horizon by adapting the Killing vector finding
methods described in \cite{Dreyer:2002mx} or \cite{Cook:2007wr}.  In
both cases it is not clear what the correct normalization of $\xi^a$
should be.  If $E$ is the quasi-local horizon energy corresponding to
the particular time translation vector field used \cite{Ashtekar00b},
then it is not clear if there is a sense in which $(E,P_\xi)$ can be
viewed as a bona fide energy-momentum four vector; for example, can we
prove $E^2=P^2_\xi+M^2$ where $M$ is the horizon mass? Some of these
issues could be studied using approximate initial data of the kind
constructed in \cite{Dennison:2006nq}.

Further study of these issues will be left to future work and in this
paper, we will work with $\xi^a$ being one of the coordinate basis
vectors in the coordinate system used in the numerical simulation;
this will yield the three components of linear momentum.  This is
obviously gauge dependent, but we shall show that there exist suitable
gauge choices in which the linear momentum passes basic but
non-trivial consistency checks.  The situation is similar to what was
observed for quasi-local horizon angular momentum
in~\cite{Campanelli:2006fy}. The true gauge invariant angular momentum
requires an accurate calculation of the axial-symmetry vector field
$\varphi^a$ on the horizon.  However, in some cases it \emph{might} be
possible to calculate the components of angular momentum by using the
rotational vector fields constructed from the coordinates used in the
simulation; but this is by no means guaranteed.  The same situation
presumably holds for linear momentum as well.

\section{Numerical Techniques and results}\label{sec:results}

We use the puncture approach~\cite{Brandt97b} along with the {\sc
TwoPunctures}~\cite{Ansorg:2004ds} thorn to compute initial data.

We confirmed that Eq.~(\ref{eq:Pxi}) yields an accurate evaluation of the
horizon momentum on the initial slice for Bowen-York data. In our tests
we found agreement between the momentum parameters and the calculated momentum
to better than 1 part in $10^6$ both for single boosted black holes with
momenta in the range $0 \leq P/M \leq 10$ and for orbiting black-hole-binary
datasets with coordinate separations as small as $0.25M$ and spins as
large at $a/m = 0.84$.

 We evolve these black-hole-binary
datasets using the {\sc LazEv}~\cite{Zlochower:2005bj} implementation
of the `moving puncture approach' which was independently proposed
in~\cite{Campanelli:2005dd, Baker:2005vv}.
 In our
version of the moving puncture approach~\cite{Campanelli:2005dd} we
replace the BSSN~\cite{Nakamura87,Shibata95, Baumgarte99} conformal
exponent $\phi$, which has logarithmic singularities at the punctures,
with the initially $C^4$ field $\chi = \exp(-4\phi)$.  This new
variable, along with the other BSSN variables, will remain finite
provided that one uses a suitable choice for the gauge.
We obtained accurate, convergent waveforms by evolving this system in
conjunction with a modified 1+log lapse, a modified Gamma-driver shift
condition~\cite{Alcubierre02a,Campanelli:2005dd}, and an initial lapse
set to $\alpha=2 \chi / (1 + \chi)$.  The lapse and shift are evolved with
\begin{eqnarray}
  (\partial_t - \beta^i \partial_i) \alpha &=& - 2 \alpha K, \label{eq:lapseeq} \\
  \partial_t \beta^a = B^a, &&
   \quad \partial_t B^a = 3/4 \partial_t \tilde \Gamma^a - \eta B^a.
   \label{eq:shifteq}
\end{eqnarray}
In this paper we will explore how the value of the gauge parameter $\eta$ affects the
quasi-local calculation of the momentum.
 We use the
{\sc Carpet}~\cite{Schnetter-etal-03b} mesh refinement driver to
provide a `moving boxes' style mesh refinement. In this approach
refined grids of fixed size are arranged about the coordinate centers
of both holes.  The {\sc Carpet} code then moves these fine grids about the
computational domain by following the trajectories of the two black
holes.
We track the location of the apparent horizons using the
{\sc AHFinderDirect} thorn~\cite{Thornburg2003:AH-finding}.

For this calculation we take $\xi^a$ in Eq.~(\ref{eq:Pxi}) to be
$(1,0,0)$, $(0,1,0)$, and $(0,0,1)$. We make this choice regardless of
slicing. This choice works for the conformally flat Bowen and York
initial data as noted above.

\subsection{Orbital Linear Momentum}

We consider initially non-spinning, unequal-mass,  orbiting black-hole
binaries, with mass ratio $3:8$, starting from an initial separation
that leads to at least two orbits prior to merger. We apply our
formula~(\ref{eq:Pxi}) to measure the linear momentum of each hole as
they orbit each other. The initial data parameters for this
configuration are summarized in Table~\ref{table:q38}. We evolved this
dataset using 10 levels of refinement with the coarsest resolution of
$h=6.4M$ and outer boundary at $320M$, and finest resolution of
$h=M/80$. In addition we evolved the same dataset after refining the
resolution at each level by a factor of $4/5$ and $4/6$.

\begin{table} 
\caption{Initial data parameters for quasi-circular orbital frequency
$\omega/M=0.05$ according to 3PN. In this run we take the
horizon mass ratio of the holes $q=m_2^H/m_1^H=3/8$, and vanishing spins, $S_i$.
The punctures are located along the $x$-axis with momenta $\vec P $
along the $y$-axis. Puncture mass
parameters are denoted by $m_{i}^p$, and horizon mass by $m_{i}^H$.} 
\begin{ruledtabular} 
\begin{tabular}{llllll} 
$x_1/M$     &1.7604572 & $m_{1}^p/M$    &0.718534207968\\
$x_2/$      &-4.7455652& $m_{2}^p/M$    &0.257487827988\\
$S_1^z/M^2$ &0.0000000&  $m_{1}^H/M$     &0.735380191\\
$S_2^z/M^2$ &0.0000000&  $m_{2}^H/M$     &0.27582649\\
$P/M$       &0.10682112& $M_{\rm ADM}/M$ &1.00001\\
$L^z/M^2$   &0.69498063& $J/M^2$         &0.69498063\\
\end{tabular} \label{table:q38}  
\end{ruledtabular}
\end{table}

\begin{figure}
\begin{center}
\includegraphics[width=2.8in,height=1.8in]{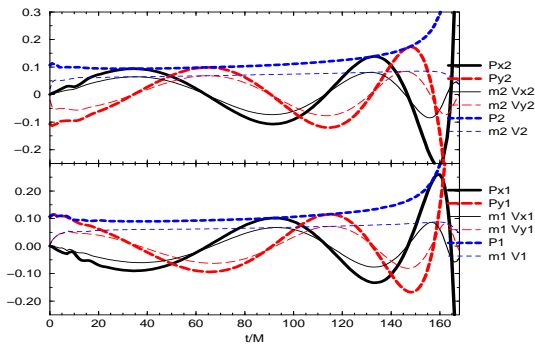}
\caption{The $x$ and $y$ components and magnitude of the individual
horizon momentum computed for the unequal mass black-hole binary with
 mass ratio $q=3/8$ until merger at $t\sim164M$.
Here $P$ denotes the quasi-local momentum and $\vec V = d \vec y/dt$.}
\label{fig:p_v_q38}
\end{center}
\end{figure}

We plot the $x$ and $y$ components of the linear momentum of each
horizon, as well as the Euclidean norm, versus time in
Fig.~\ref{fig:p_v_q38}. We have calculated the momentum both using
formula~(\ref{eq:Pxi}) and the purely coordinate momentum
$\vec P_i = m_i d \vec y_i/dt$, where $\vec y_i$ is the trajectory
of puncture $i$
and $m_i$ is the horizon mass.  Note that the coordinate momentum is
initially zero due to our choice $\beta^i(t=0) = B^i(t=0) = 0$,
and that it is consistently lower than the initial momentum of the
holes (and decreases at late-times). Therefore, we expect that the quasi-local momentum provides a
more accurate measurement than the coordinate momentum. Furthermore, we
expect that the momentum will increase as the binary inspirals, in
qualitative agreement with the behavior of the quasi-local momentum.
Note that both evaluations of the momentum agree in phase, hence we
expect that the quasi-local momentum, which has a more accurate amplitude,
will provide a more accurate measurement of the instantaneous angular velocity.

\subsection{Recoil momentum}

In this section we study the recoil velocity of an already
merged black-hole binary that acquires linear momentum as a
reaction to the asymmetric radiation of gravitational waves during
the inspiral and merger phases. We apply formula~(\ref{eq:Pxi}) to
a single, slightly distorted, black hole and we attempt to compute
speeds of the order $v/c \sim 0.0001$, which is three orders of magnitude
smaller than the speeds in the orbital case.

We evolve two equal-mass black holes with opposite spins (pointing
perpendicular to the line joining the holes), starting
from rest, at an intermediate separation.
The initial data parameters are summarized in Table~\ref{table:headon}.
For this configuration the recoil momentum, as determined by an
extrapolation of $\psi_4$ to $r=\infty$, is $P_{y}/M=(20.4\pm0.5)\KMS$
($P_x = P_z = 0$).
We evolved these data using 9 levels of refinement with coarsest
resolution $h=8M$ and finest resolution $h=M/32$, as well  higher
resolutions runs with finest resolutions of $M/48$ and $M/64$ (with a 
corresponding increase in gridpoints on all levels). Here we explore
the dependence of the measured kick both on resolution and the $\eta$
parameter in Eq.~(\ref{eq:shifteq}).
(Note that Eq.~(\ref{eq:lapseeq}) approximates maximal slicing, i.e.\ 
$K\to0$, at late times.)

\begin{table} 
\caption{Initial data parameters for the equal mass opposite spinning 
black holes.  The punctures are 
located along the $x$-axis starting from rest.
Spins are denoted by $S_i$, puncture mass
parameters by $m_{i}^p$, and horizon mass by $m_{i}^H$.} 
\begin{ruledtabular} 
\begin{tabular}{llllll} 
$x_1/M$     &-3.5000 & $m_{1}^p/M$    &0.427644\\
$x_2/M$     &+3.5000 & $m_{2}^p/M$    &0.427644\\
$S_1^z/M^2$ &+0.15   & $m_{1}^H/M$    &0.517407\\
$S_2^z/M^2$ &-0.15   & $m_{2}^H/M$    &0.517407\\
$P/M$       & 0.0000 & $M_{\rm ADM}/M$ &0.999998\\
$L^z/M^2$   & 0.0000 & $J/M^2$         &0.0000  \\
\end{tabular} \label{table:headon}  
\end{ruledtabular}
\end{table}

\begin{figure}
\begin{center}
\includegraphics[width=1.68in]{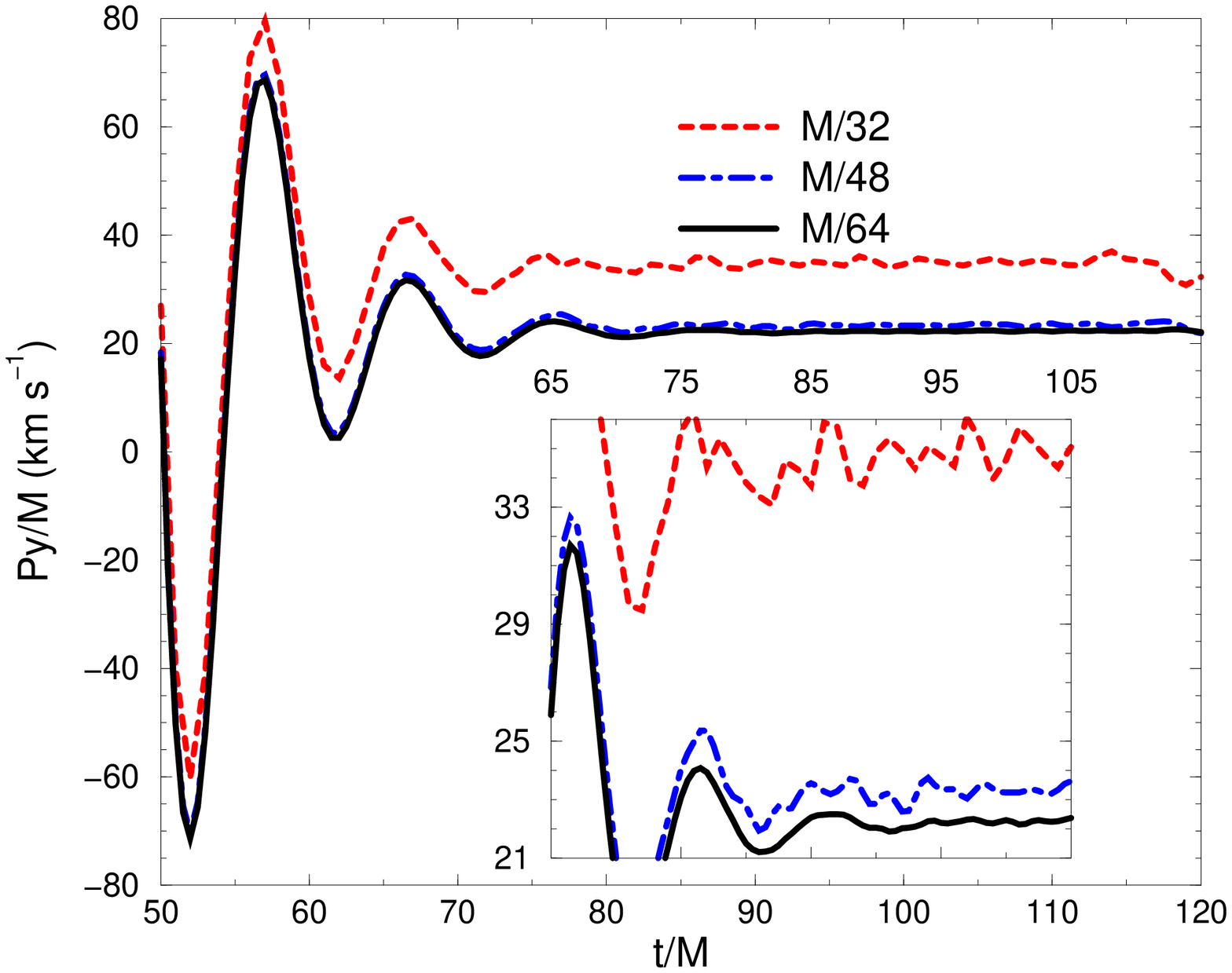}
\includegraphics[width=1.68in]{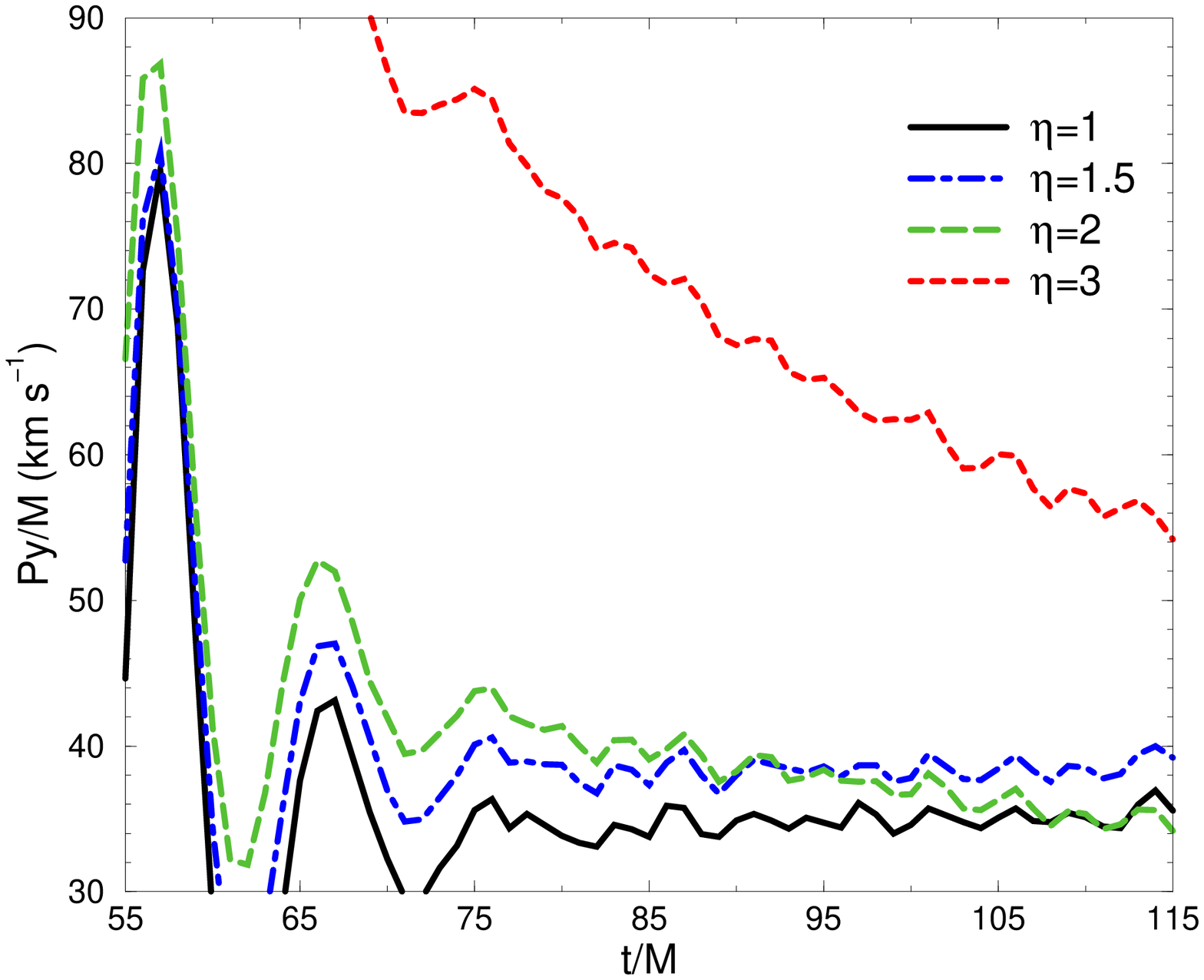}
\caption{The remnant recoil velocity for the head-on test as determined
by the isolated horizon formula. The panel on the left shows $P_y$
versus resolution for $\eta=1$. Note the rapid convergence of the asymptotic plateau. The panel on the right shows
$P_y$ versus $\eta$ for $h=M/32$.
The expected
value of the recoil is $P_{y}/M=(20.4\pm0.5)\KMS$.}
\label{fig:headon_mom_vres_veta}
\end{center}
\end{figure}

The left panel of Fig.~\ref{fig:headon_mom_vres_veta} shows the quasi-local recoil momentum
versus time for $\eta=1$ and the three resolutions. Note the rapid
convergence of the asymptotic value of the momentum versus resolution.
The measured convergence rate is greater than 5, but even at a central
resolution of $M/48$ the error in the recoil is within $3 \KMS$.
An extrapolation to infinite resolution gives $P_{y}/M = (21.5\pm0.5)\KMS$.
The small disagreement between the quasi-local recoil and recoil calculated
from $\psi_4$ is due to the use of a non-vanishing  $\eta$.
The right panel of Fig.~\ref{fig:headon_mom_vres_veta} shows the dependence of the
quasi-local momentum on the gauge via variations in $\eta$. Note that
while $\eta$ has essentially no effect on the recoil calculated from
$\psi_4$, the distortion in the gauge caused by large $\eta$ has a
significant affect on the quasi-local momentum for $\eta > 2$. In
particular, for large $\eta$, the quasi-local momentum displays a
slow secular decay towards
a final asymptotic value, with  both an increasing amplitude and
decreasing decay rate, as $\eta$ is increased.

Note that the quasi-local formula provides an accurate
measurement of the recoil at $t\sim 80 M$, while a measurement based on
$\psi_4$ requires evolutions of at least $150M$ for observers at $40M$.

\begin{figure}
\begin{center}
\includegraphics[width=2.8in,height=1.8in]{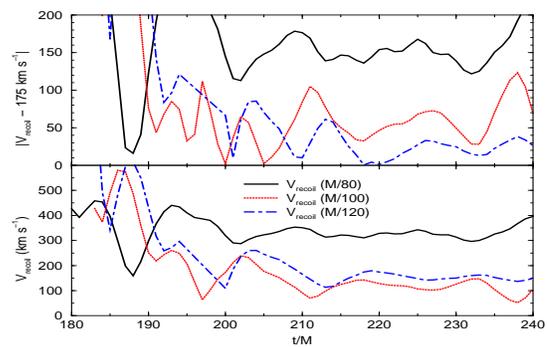}
\caption{The recoil velocity of the remnant (bottom) and
error (top) for the orbiting binary versus resolution for $\eta=2$.}
\label{fig:q38_kick}
\end{center}
\end{figure}

Figure~\ref{fig:q38_kick} shows the quasi-local recoil momentum calculation
for the orbiting binary configuration as a function of time
and resolution. The expected recoil velocity in this case is $V=175\KMS$.

\section{Conclusion}\label{sec:conclusion}

There is currently no other fully relativistic method to compute the
linear momentum of black holes in a binary. Knowledge of the linear
momentum of each hole provides an important diagnostic in comparing
fully-nonlinear results with post-Newtonian predictions of the
trajectory and waveform (for instance as an alternative measure of the
instantaneous angular velocity of the binary system).

The quasi-local approach we propose represents not only an alternative
measure of the linear momentum, but also provides an accurate
measurement of the recoil much sooner than can be obtained from the
waveform.

In future work we will study a more coordinate independent and robust
derivation of isolated horizon inspired formulae to evaluate the
linear momentum of black holes. This can be achieved by an improved
evaluation of the $\xi^a$ in addition to extrapolation to the
$\eta\to0$ and $h\to0$ limits, and making greater use of the horizon
geometry than was done here.

\begin{acknowledgments} 
  We thank Erik Schnetter for technical support and for providing
  CARPET, and Marcus Ansorg for providing the {\sc TwoPunctures}
  initial data thorn.  We are also very grateful to Manuela Campanelli,
  Abhay Ashtekar, and
  Sergio Dain for valuable discussions and suggestions.  We gratefully
  acknowledge NSF for financial support from grant PHY-0722315.
  Computational resources were provided by the Lonestar cluster at
  TACC, the Funes cluster at UTB, and the NewHorizons cluster at RIT.
\end{acknowledgments} 

\bibliographystyle{apsrev}
\bibliography{../bibtex/references}

\end{document}